\documentclass[
letterpaper,
twoside,
twocolumn,
superscriptaddress,
amsmath,amssymb,
prb,
floatfix,
]{revtex4-2}

\usepackage{graphicx}% Include figure files
\usepackage{dcolumn}% Align table columns on decimal point
\usepackage{bm}% bold math
\usepackage{hyperref}% add hypertext capabilities
\usepackage{xr-hyper}
\usepackage{textcomp}
\usepackage[mathlines]{lineno}% Enable numbering of text and display math
\usepackage[dvipsnames]{xcolor}
\begin{document}

\title{Anisotropy and effective medium approach in the optical response of 2D material heterostructures}% Force line breaks with \\

\author{Bruno Majérus}
\affiliation{Laboratoire de physique du solide (LPS) \& Namur Institute of Structured Matters (NISM), University of Namur, 61 rue de Bruxelles, B-5000 Namur, Belgium.}
\author{Emerick Guillaume}
\affiliation{Laboratoire de physique du solide (LPS) \& Namur Institute of Structured Matters (NISM), University of Namur, 61 rue de Bruxelles, B-5000 Namur, Belgium.}
\affiliation{IMOMEC, IMEC vzw, Wetenschapspark 1, 3590 Diepenbeek, Belgium}
\affiliation{UHasselt, Institute for Materials Research (IMO-IMOMEC), Agoralaan, 3590 Diepenbeek, Belgium}
\author{Pascal Kockaert}
\affiliation{OPERA-photonics, Université libre de Bruxelles (U.L.B.), 50 Avenue F. D. Roosevelt, CP 194/5, B-1050 Bruxelles, Belgium}
\author{Luc Henrard}
\affiliation{Laboratoire de physique du solide (LPS) \& Namur Institute of Structured Matters (NISM), University of Namur, 61 rue de Bruxelles, B-5000 Namur, Belgium.}

\begin{abstract}

2D materials offer a large variety of optical properties, from transparency to plasmonic excitation. They can be structured and combined to form heterostructures that expand the realm of possibility to manipulate light interactions at the nanoscale. Appropriate and numerically efficient models accounting for the high intrinsic anisotropy of 2D materials and heterostructures are needed. In this article, we retrieve the relevant intrinsic parameters that describe the optical response of a homogeneous 2D material from a microscopic approach. Well-known effective models for vertical heterostructure (stacking of different layers) are retrieved. We found that the effective optical response model of horizontal heterostructures (alternating nano-ribbons) depends of the thickness. In the thin layer model, well adapted for 2D materials, a counter-intuitive in-plane isotropic behavior is predicted. We confront the effective model formulation with exact reference calculations such as \textit{ab-initio} calculations for graphene, hexagonal boron nitride (hBN), as well as corrugated graphene with larger thickness but also with classical electrodynamics calculations that exactly account for the lateral structuration. 
\end{abstract}

%\keywords{Suggested keywords}%Use showkeys class option if keyword
                              %display desired
\maketitle

%\tableofcontents

\section{Introduction}

 The extraordinary optical and electromagnetic (EM) properties of two-dimensional (2D) materials have been broadly investigated from visible light to microwaves \cite{Rouhi2012,Grigorenko2012,Bozzi2015,Low2014}, leading to developments in various domains such as photovoltaics \cite{Park2012,Das2019}, biosensors \cite{Justino2017,Chattopadhyay2015}, superabsorbers \cite{Batrakov2014,Lobet2016a} and transparent conducting films \cite{Kasry2010,Park2012}. 
 The description of the EM response of a single layer has been debated recently \cite{Tancogne-Dejean2015, Matthes2016, Merano2015, Jayaswal2018,Li2018, Majerus2018,Guilhon2019, Xu2021,DellAnna2022a,Majerus2023} based on a thin film model which assigns an effective permittivity to a layer with a given thickness, or on a 2D model which sets a surface susceptibility or conductivity at the interface between two media \cite{Majerus2018,Li2018,Xu2021}. 

It is clear that anisotropy is essential in the description of 2D materials. As an example, recent ellipsometry results on MoS\textsubscript2 and graphene have shown that the out-of-plane response plays a crucial role in their optical response~\cite{Jayaswal2018,Xu2021}.  In particular, the comparison between the thin film and surface susceptibility models required this out-of-plane response to be carefully handled as the materials are not periodic in that direction ~\cite{Majerus2018,Majerus2023}. 

The stacking of 2D layers modifies their electronic properties as shown for the stacking order of multilayer graphene \cite{Latil2006,Cao2018,Yelgel2016,Hagymasi2022} or for the transition from direct to indirect band gap for TMDs \cite{Mann2014,Kadantsev2012}. These effects mainly occur close to the Fermi level and affect less the EM response in the visible or U.V. range. The stacking also results in long-range (electrostatic) interactions that modify the optical response in the absence of change in the electronic structure of the system \cite{Tancogne-Dejean2015,Majerus2023}. This long-range interaction should also be accounted for to retrieve the single-layer optical response from quantum simulation based on supercell techniques (periodic repetition of the single-layer separated with vacuum)~\cite{Tancogne-Dejean2015}. 

The high number of possible heterostructures and their atomic complexity, as well as their intrinsic anisotropy, demand a robust but computationally tractable approach.
Vertical heterostructures, the stacking of identical or different 2D layers, have been widely investigated in the last decade \cite{Poddubny2013,Bludov2013,Batrakov2014,Novoselov2016,Li2017a,Wang2017,Farmani2017,Ren2019,Li2020,Zhang2021}, in particular with an effective medium approach \cite{Poddubny2013,Chebykin2012, Majerus2018}. However, the thickness range of validity for a thin film or surface susceptibility effective models has not been explored. On the other hand, horizontal heterostructures of single-layer materials have been less studied \cite{Christensen2012,Li2017a,Li2020,Das2018} and the effective models have not been confronted to exact methods that account for the structuring. 

In this paper, first, we investigate the validity of the effective model for vertical 2D heterostructures with a graphene-hBN bilayer and we explore the limits of this approach with respect to the number of layers. Second, turning to horizontal heterostructure, we show that the optical response of 2D materials nanoribbons are correctly described with a different effective model than thick ribbons and nanorods~\cite{Li2020,Correas-Serrano2015}. In particular, we show that 2D materials horizontal heterostructures optically behave like a uniform uniaxial material, isotropic in the plane.

In section \ref{sec:theory}, starting from a microscopic framework, we define surface susceptibilities of 2D materials which are independent of the thickness, both for in-plane and out-of-plane polarisation, retrieving in a microscopic approach, that includes among others smooth transitions between layers, the results obtained from a macroscopic point of view in  ~\cite{Majerus2023}. On this basis, the effective models for vertical and horizontal heterostructures are retrieved. The reference numerical methods, with which effective model will be compared, are described in section \ref{sec:Methods}. First-principle quantum approach (time-dependent density functional theory - TDDFT) has been performed when the size of the system permits, e.g. for vertical heterostructure, and classical electrodynamics approach that exactly accounts for the structuring (Rigorous coupled wave analysis (RCWA)) has been used for horizontal heterostructure. In section \ref{sec:Results} we apply the effective models on various heterostructures and we compare the results with reference simulations. Graphene multilayers and a graphene-hBN bilayer are investigated for vertical heterostructures while graphene nanoribbons and graphene-hBN hetero-nanoribbons are studied for horizontal heterostructures.

\section{Effective Medium theory for 2D materials and heterostructures}\label{sec:theory}
In this section, macroscopic response functions are related to microscopic response functions and effective models are deduced for heterostructures.

\subsection{The irreducible and external susceptibilities}
 The time dependencies of the potentials are supposed to be harmonic, i.e. $V(t)\propto e^{i\omega t}$. Below, this factor is not shown for the sake of conciseness.
 $V_{app}\left(\mathbf{r}\right)$ is a periodic potential with a uniform amplitude $\tilde{V}_{app}$ applied on a material at a microscopic (atomic) scale,
\begin{equation}
        V_{app}\left(\mathbf{r}\right)=\tilde{V}_{app}\,e^{i\mathbf{k}\cdot\mathbf{r}}.
\end{equation}
The induced and total potentials $V_{ind}\left(\mathbf{r}\right)$ and $V_{tot}\left(\mathbf{r}\right)$ can be written as follows:
\begin{eqnarray}
    V_{ind}\left(\mathbf{r}\right)&=&\tilde{V}_{ind}\left(\mathbf{r}\right)e^{i\mathbf{k}\cdot\mathbf{r}},\\
    V_{tot}\left(\mathbf{r}\right)&=&\tilde{V}_{tot}\left(\mathbf{r}\right)e^{i\mathbf{k}\cdot\mathbf{r}},
\end{eqnarray}
where the spatial variation of the functions $\tilde{V}_{tot}(\mathbf{r})$ and $\tilde{V}_{ind}(\mathbf{r})$ are related to the local fields (LF) and have the same spatial periodicity as the unit cell of the material. 
We define here the irreducible susceptibility $\chi\left(\mathbf{r},\mathbf{r}'\right)$ and the external susceptibility  $\xi\left(\mathbf{r},\mathbf{r}'\right)$ by:
\begin{align}
         \tilde{V}_{ind}\left(\mathbf{r}\right) &=-\int{\chi\left(\mathbf{r},\mathbf{r}'\right)\tilde{V}_{tot}\left(\mathbf{r}'\right)}\,d^{3}\mathbf{r}' \label{eq:chi_micro},\\ 
     \tilde{V}_{ind}\left(\mathbf{r}\right) &=-\int{\xi\left(\mathbf{r},\mathbf{r}'\right)\tilde{V}_{app}}\,d^{3}\mathbf{r}'.
     \label{eq:xi_micro}
\end{align}
where the integrals span over the whole space.
These susceptibilities can be calculated from the more usual irreducible and external polarizabilities~\cite{SupplementalMaterial,Bernadotte2013}. The macroscopic dielectric function of a material is obtained through the average of the potential over the unit cell $\Omega$  ~\cite{Wiser1963,Huser2013a,Hybertsen1987} :
\begin{equation}
\varepsilon_{M}=\frac{\tilde{V}_{app}}{\left\langle \tilde{V}_{tot}\left(\mathbf{r}\right)\right\rangle }.
\label{eq:epsm}
\end{equation}

Taking the average of the total potential from eq. (\ref{eq:xi_micro}) using the fact that the total field is the sum of the applied and the induced fields, it comes:
\begin{align}
\frac{1}{\varepsilon_{M}}&=1-\frac{1}{V}\int_\Omega\int \xi\left(\mathbf{r},\mathbf{r}'\right)d^{3}\mathbf{r}'d^{3}\mathbf{r},\\
\frac{1}{\varepsilon_{M}}&\equiv 1-\xi_{M},
\label{eq:inv_epsm}
\end{align}
with $V$ the volume of the unit cell. The macroscopic external susceptibility $\xi_{M}$ defined by eq.~(\ref{eq:inv_epsm}) relates the total displacement field to the polarization field
\begin{equation}
    \mathbf{P}=\xi_{M}\mathbf{D}.
    \label{eq:xi_m}
\end{equation}
This susceptibility has already been defined in \cite{Majerus2023} as the displacement susceptibility. In contrast, the macroscopic irreducible susceptibility $\chi_M$, the usual susceptibility of electromagnetism, is defined by
\begin{equation}
    \mathbf{P}=\varepsilon_0\chi_{M}\mathbf{E},
\end{equation}
and is obtained from eq.~(\ref{eq:inv_epsm}),
\begin{equation}
\chi_M= 1- \varepsilon_M = \frac{\xi_{M}}{1-\xi_{M}}.
\label{eq:chim_xim}
\end{equation}

In general $\chi_M$ cannot be obtained directly by averaging eq.~(\ref{eq:chi_micro}), because of the spatial variations of the total field $\tilde{V}_{tot}$ in the unit cell, associated with the local fields effect. However, if the material is homogeneous, the LF are negligible and the total field can be replaced by its spatial average in eq.~(\ref{eq:chi_micro}) \cite{Wiser1963,Yan2011} and
\begin{align}
\varepsilon_{M}&=1+\frac{1}{V}\int_{\Omega}\int \chi\left(\mathbf{r},\mathbf{r}'\right)d^{3}\mathbf{r}'d^{3}\mathbf{r}\label{eq:epsm_chi},\\
\varepsilon_{M}&\equiv1+\chi_{M},
\end{align}
where $\chi_{M}$ is equal to that obtained using eq.~(\ref{eq:chim_xim}).

\begin{figure}    \centering
    \includegraphics[width=\linewidth, trim = 1cm 6cm 7cm 4cm, clip]{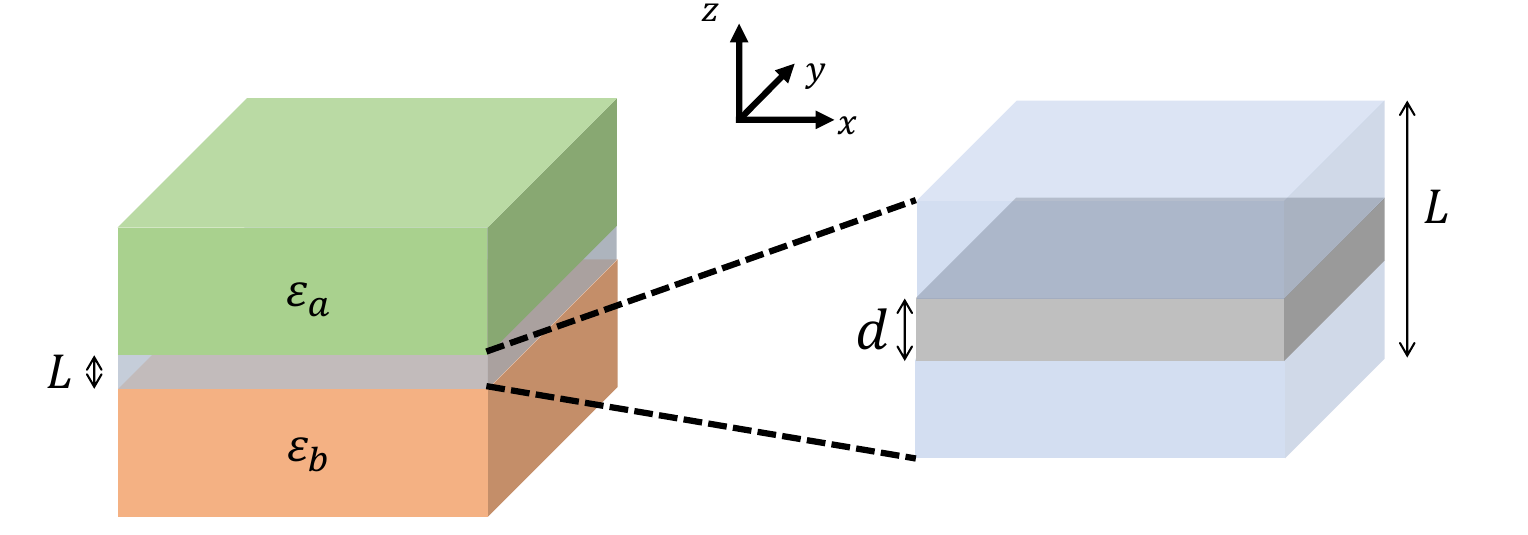}
    \caption{Model of a 2D material layer between two media. }
    \label{Fig:model_layer}
\end{figure}

\subsection{Permittivity and surface susceptibilities of 2D materials}
\begin{figure}[b]
    \centering
    \includegraphics[width=\linewidth]{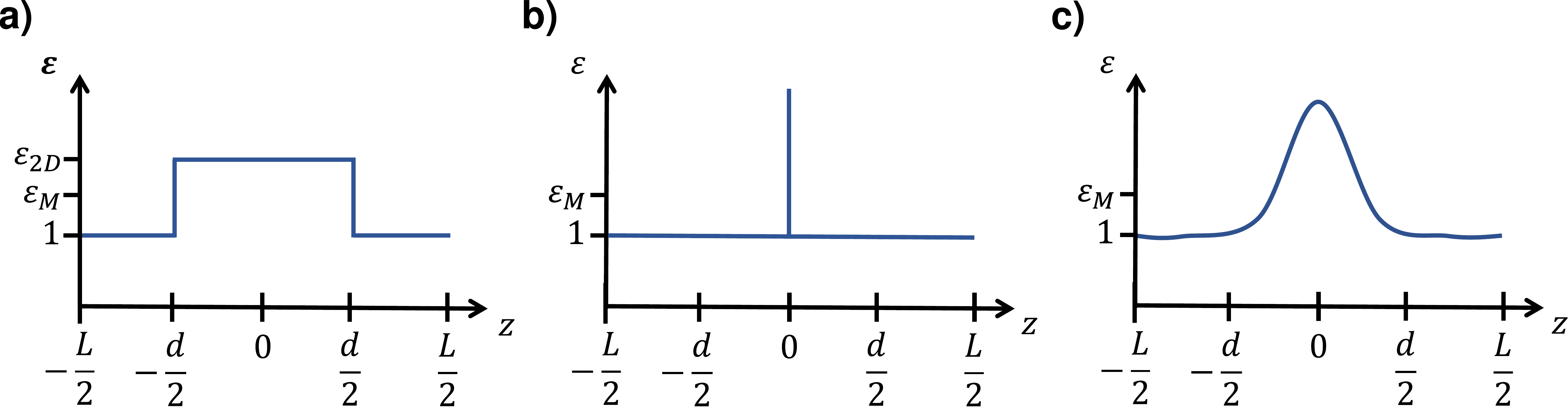}
    \caption{Permittivity of a 2D material as a function of the position across the layer within: (a)~a thin-film model; (b)~a surface polarization model; (c)~a more general description of the spatial variation.}
    \label{Fig:model_perm}
\end{figure}

Bidimensional materials cannot be considered strictly 2D because the electronic wave function extends in the normal direction. Therefore, the microscopic dielectric function varies along this direction. In order to determine this permittivity both numerically (e.g. using TDDFT \cite{Tancogne-Dejean2015,Huser2013a}) and experimentally (e.g. using ellipsometry \cite{Majerus2018a,Jayaswal2018,Xu2021}), 2D materials can be modelled as a layer with a constant permittivity over a thickness $L$ \cite{Tancogne-Dejean2015,Huser2013a}. In the most general case, the layer of thickness $L$ is embedded between two media of different permittivities $\varepsilon_a$ and $\varepsilon_b$ as represented in fig.~\ref{Fig:model_layer} (left). Following~\cite{Majerus2023}, this layer can be described as a 2D material of permittivity $\varepsilon_{2D}$ with thickness $d$ surrounded by vacuum (fig.~\ref{Fig:model_layer}, right). This vacuum may for example represent the vacuum layers of the supercell used in DFT, or the interlayer distance with another 2D material in the case of heterostructures.
The spatial variation of the permittivity in this layered system (vacuum -- 2D material -- vacuum) is schematically represented in fig.~\ref{Fig:model_perm}. When \(d=L\), it corresponds to the thin film model. When \(d\rightarrow0\),  the 2D material is infinitely thin and the permittivity is represented using a Dirac distribution as for a finite surface polarization at the interface between two materials. These two approaches have been formally combined in~\cite{Majerus2023}. Other models can be imagined such as a continuous permittivity, with a maximum value at the center of the atomic layer as represented in fig.~\ref{Fig:model_perm}c. However, we show in the following that at a macroscopic scale, all these descriptions are equivalent. 

To rigorously define the surface susceptibilities of 2D materials, we use the effective model describing the three-layer system (vacuum -- 2D material -- vacuum) represented on the right side of fig.~\ref{Fig:model_layer} and in fig.~\ref{Fig:model_perm}a. This effective model comes, for in-plane polarization, from the conservation over the whole system of the tangential component of the electric field, in which case we can use the parallel capacitors equation, eq.~(S9). For out-of-plane polarization, it comes from the conservation of the normal component of the displacement field, in which case we can use the series capacitors equation eq. (S12) \cite{SupplementalMaterial,Bechstedt1986,Mohammadi2005}. For in-plane polarization, it gives:
\begin{equation}
\varepsilon_{eff}^{\parallel}=\frac{L-d}{L}\varepsilon_{vac}+\frac{d}{L}\varepsilon_{2D}^{\parallel},
\label{eq:eps_parall}
\end{equation}
and for out-of-plane polarization 
\begin{equation}
\frac{1}{\varepsilon_{eff}^{\perp}}=\frac{L-d}{L}\frac{1}{\varepsilon_{vac}}+\frac{d}{L}\frac{1}{\varepsilon_{2D}^{\perp}},
\end{equation}
The permittivities $\varepsilon_{2D}^{\parallel}$ and $\varepsilon_{2D}^{\perp}$ are the permittivities of the layer of thickness $d$ for in-plane and out-of-planes polarizations respectively.

The fact that the effective permittivity modelling is different for $\parallel$ and $\perp$ directions is related to the role of the local field on the optical properties of stratified media and 2D materials. LF affects mostly fields polarized perpendicularly to the sheets, while it may be neglected for in-plane polarization \cite{Marinopoulos2002a,Tancogne-Dejean2015}. Since LFs are negligible for this polarization, $\varepsilon_{M}$ of eq.~(\ref{eq:epsm_chi}) can be used to evaluate $\varepsilon_{2D}^{\parallel}$ in eq. (\ref{eq:eps_parall}) and, with $\varepsilon_{vac}=1$, we obtain
\begin{equation}
\varepsilon_{eff}^{\parallel}=1+\frac{1}{L\,S}\int_{\Omega}\int \chi^{\parallel}\left(\mathbf{r},\mathbf{r}'\right)d^{3}\mathbf{r}'d^{3}\mathbf{r},
\label{eq:eps-eff_parall}
\end{equation}
with $S=V/d$ the surface of the unit cell. 

We can define the surface irreducible susceptibility for in-plane polarization of a 2D material as
\begin{equation}
        \chi_{S}^{\parallel}=\frac{1}{S}\int_{\Omega}\int \chi^{\parallel}\left(\mathbf{r},\mathbf{r}'\right)d^{3}\mathbf{r}'d^{3}\mathbf{r}  \\
        \label{eq:surf_chi}
\end{equation}
such that
\begin{equation}
\varepsilon_{eff}^{\parallel}=1+\frac{\chi_{S}^{\parallel}}{L}.
\label{eq:eps_eff_parall2}
\end{equation}
The second term of the right-hand side of eq.~(\ref{eq:eps-eff_parall}) is the average value of the microscopic susceptibility $\chi^{\parallel}\left(\mathbf{r},\mathbf{r}'\right)$ over a surface $S$ and a height $L$. Therefore, it does not depend directly on the variation profile of the permittivity in the layer of thickness L or on the distance $d$ and the equation (\ref{eq:eps_eff_parall2}) is valid for other models such as those represented in fig.~\ref{Fig:model_perm}b and c. While the in-plane surface susceptibility \(\chi_{S}^{\parallel}\) is independent of the chosen thickness $L$, as it accounts only for the response of the 2D material in the volume of thickness $d$, we note that the effective permittivity $\varepsilon_{eff}^{\parallel}$ depends on $L$. 

The same reasoning can be performed for the out-of-plane polarization. Because the LF cannot be neglected, eq.~(\ref{eq:inv_epsm}) is used to obtain the effective permittivity of the layer
\begin{equation}
\frac{1}{\varepsilon_{eff}^{\perp}}=1-\frac{1}{L}\frac{1}{S}\int_{\Omega}\int \xi^{\perp}\left(\mathbf{r},\mathbf{r}'\right)d^{3}\mathbf{r}'d^{3}\mathbf{r}.
\end{equation}
As before, the second term of the right-hand side of eq.~(\ref{eq:eps-eff_parall}) is the average value of the microscopic susceptibility $\xi^{\perp}\left(\mathbf{r},\mathbf{r}'\right)$ over a surface $S$ and a height $L$. The surface external susceptibility for out-of-plane polarization is then
\begin{equation}
        \xi_{S}^{\perp}=\frac{1}{S}\int_{\Omega}\int \xi^{\perp}\left(\mathbf{r},\mathbf{r}'\right)d^{3}\mathbf{r}'d^{3}\mathbf{r} \label{xi_surf},
\end{equation}
and the effective out-of-plane permittivity of the layer is
\begin{equation}
\varepsilon_{eff}^{\perp}=\frac{1}{1-\frac{\xi^{\perp}_{S}}{L}}.
\label{eq:eps_eff_perp}
\end{equation}
As for the in-plane response, the surface external susceptibility is independent of the thickness, but the effective permittivity $\varepsilon_{eff}^{\perp}$ depends on the thickness of the layer.

We emphasize, as stated above, that the surface susceptibilities of eqs. (\ref{eq:surf_chi}) and (\ref{xi_surf}) are model-independent, in the sense that they do not depend on the exact variation profile of the permittivity (see fig.~\ref{Fig:model_perm}). They describe the average response of the 2D material at this interface and they are the relevant quantities to describe the optical response of 2D materials at a macroscopic scale. In particular, they are related to the surface polarization field by
\begin{equation}
    \mathbf{P^{\parallel}_S}=\varepsilon_0\chi_S^{\parallel}\mathbf{E}^{\parallel},
    \label{eq:chi_s}
\end{equation}
and 
\begin{equation}
   P^{\perp}_S=\xi_S^{\perp}{D}^{\perp}.
    \label{eq:xi_s}
\end{equation}

Note that $\chi_S^{\perp}$ has been used to characterize the out-of-plane polarization of 2D materials ~\cite{Matthes2016,Majerus2018,Xu2021,DellAnna2022a}. However, from eq.~(\ref{eq:chim_xim}), we see that it is not an intrinsic quantity as it depends on the thickness L.
\begin{equation}
    \chi_{S}^{\perp}=\frac{\xi_{S}^{\perp}}{1-\frac{\xi_{S}^{\perp}}{L}}.
\end{equation}
Moreover, when the constitutive relation \(P_{S}^\perp = \varepsilon_0 \chi_{S}^\perp E^\perp\) is used instead of eq. (\ref{eq:xi_s}), the obtained surface response function are found to problematically depend on the surrounding media ~\cite{Matthes2016,Majerus2018,Xu2021,DellAnna2022a}. In the case of a 2D layer in vacuum or when the out-of-plane susceptibility is neglected, the derived optical responses are not affected. But recent ellipsometry measurements on graphene and MoS\textsubscript{2} in the visible range~\cite{Xu2021} cannot be interpreted with a model that neglects the out-of-plane response of the 2D material, reinforcing the need to define truly intrinsic quantities for the 2D layer, independent of the external media \cite{Majerus2023}.

The microscopic susceptibilities of 2D materials can be obtained numerically {\it ab initio} for a periodic system in a supercell approach  \cite{Hybertsen1987,Yan2011}, with a vacuum layer of a few nanometers separating repeating layers to avoid short-range interactions between them. The thickness $L$ then corresponds to the height of the supercell. If, in the {\it ab initio} codes, the macroscopic permittivity is directly given as the result of the quantum calculations, the surface susceptibilities can be derived from eq.~(\ref{eq:eps_eff_parall2}) and (\ref{eq:eps_eff_perp}). The permittivities or susceptibilities can then be used in classical electrodynamics approaches, implementing each 2D material with a 2D or 3D model.

\subsection{Effective model for 2D-material heterostructures}\label{sec:effective_models}

\begin{figure}
    \centering
    \includegraphics[width=\linewidth]{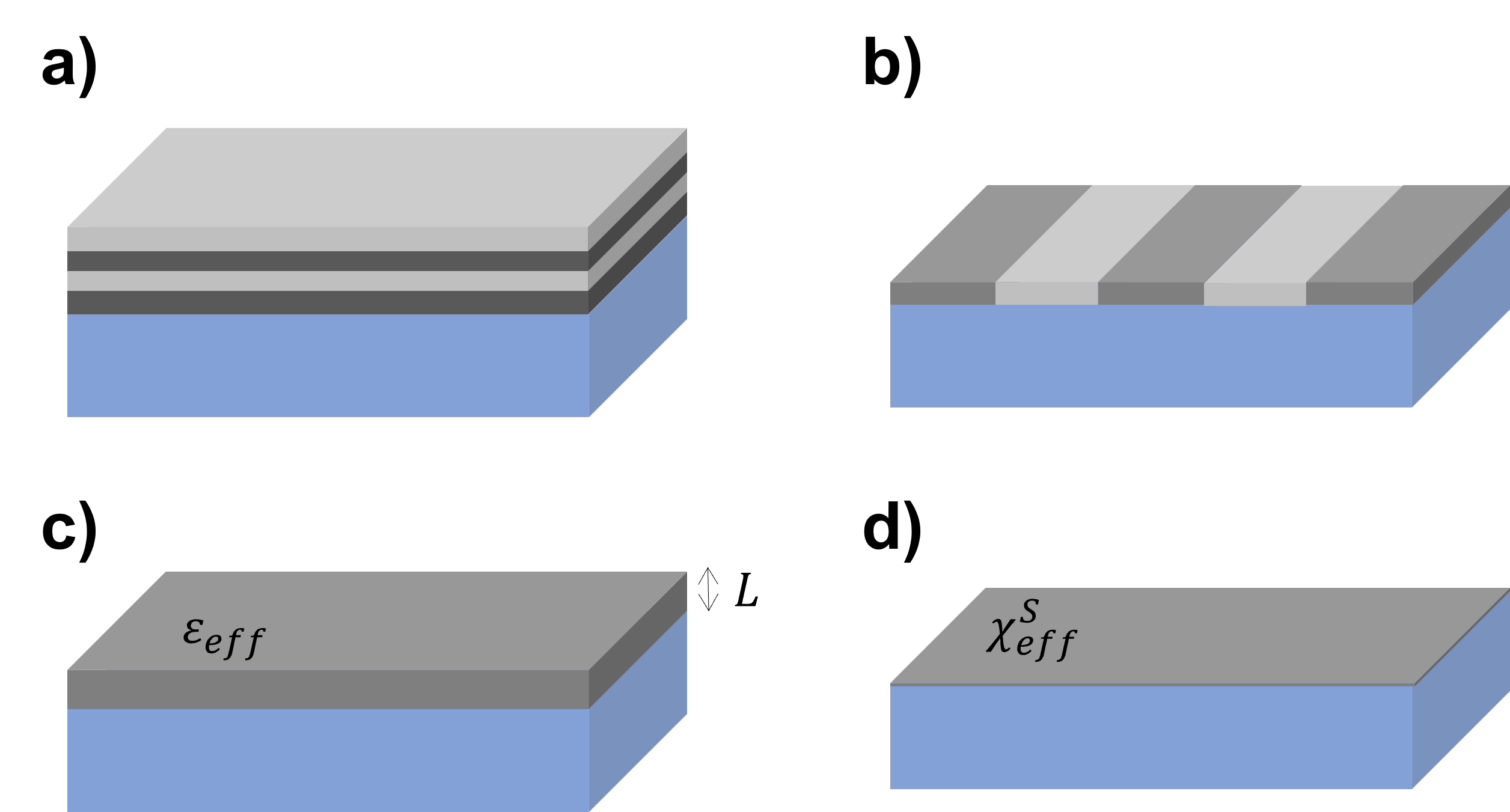}
    \caption{Schematic representation of the considered systems and the effective models. Vertical heterostructures (a) and horizontal heterostructures (b) of 2D materials can be represented by a thin film (c) or a surface polarization (d). }
    \label{Fig:schema1}
\end{figure}

In this section, the in-plane and out-of-plane surface susceptibilities of horizontal and vertical heterostructures are related to the bulk effective permittivities of a thin film of finite thickness or to the surface susceptibilities of a surface polarization as shown in fig.~\ref{Fig:schema1}, where (a) and (b) represent the heterostructures and (c) and (d) the effective models (3D or 2D).

\subsubsection{Effective model for vertical heterostructures} \label{sec:vert_het}

A vertical heterostructure is modelled here as alternating layers of 2D materials and vacuum  (fig.~\ref{Fig:schema1}a), which can be seen as a generalisation of the approach of the previous section. The effective permittivity of a multilayer can be found using the parallel capacitors equation (eq. (S9)) \cite{Poddubny2013,Chebykin2012}. Moreover, the effective surface irreducible susceptibility of a purely 2D material equivalent to the multilayer can be deduced from eq. (S9) using eq.~(\ref{eq:eps_eff_parall2}):
\begin{equation}
\chi_{S,eff}^{\parallel}=\sum_i \chi_{S,i}^{\parallel}.
\label{eq:sum_chi_s}
\end{equation}
where the sum spans on each layer indexed $i$. A further analysis of the validity of this approach is proposed in \cite{Majerus2023}.

Similarly, from the series capacitors equation (eq. (S12)) and using eq.~(\ref{eq:eps_eff_perp}), an effective surface external susceptibility for the out-of-plane polarization is obtained as:
\begin{equation}
\xi_{S,eff}^{\perp}=\sum_i \xi_{S,i}^{\perp}.
\label{eq:sum_xi_s}
\end{equation}
Equations (\ref{eq:sum_chi_s}) and (\ref{eq:sum_xi_s}) are equivalent to eqs. (53) and (56) of \cite{Majerus2023}, validating the coherence of the two approaches.

\subsubsection{Effective model for lateral heterostructures}

A horizontal heterostructure corresponds to alternating ribbons of 2D materials (fig.~\ref{Fig:schema1}b). This kind of geometry was not considered in~\cite{Majerus2023}  and no model published up to now can provide the counter-intuitive results that are obtained for thin ribbons below.

Following the same approach as for vertical heterostructure, the effective susceptibilities of thick ribbons are found, reproducing well-known results \cite{Liu2008,Li2020}:
\begin{equation}
\xi_{S,eff}^{x}=\sum f_{i}\xi_{S,i}^{x}
\label{eq:ribbon-thick-xi}
\end{equation}
\begin{equation}
\chi_{S,eff}^{y,z}=\sum f_{i}\chi_{S,i}^{y,z}.
\label{eq:ribbon-thick-chi}
\end{equation}
where $f_i$ is the volume filling fraction of each type of ribbon.

\begin{figure}
\begin{centering}
\includegraphics[width=\linewidth]{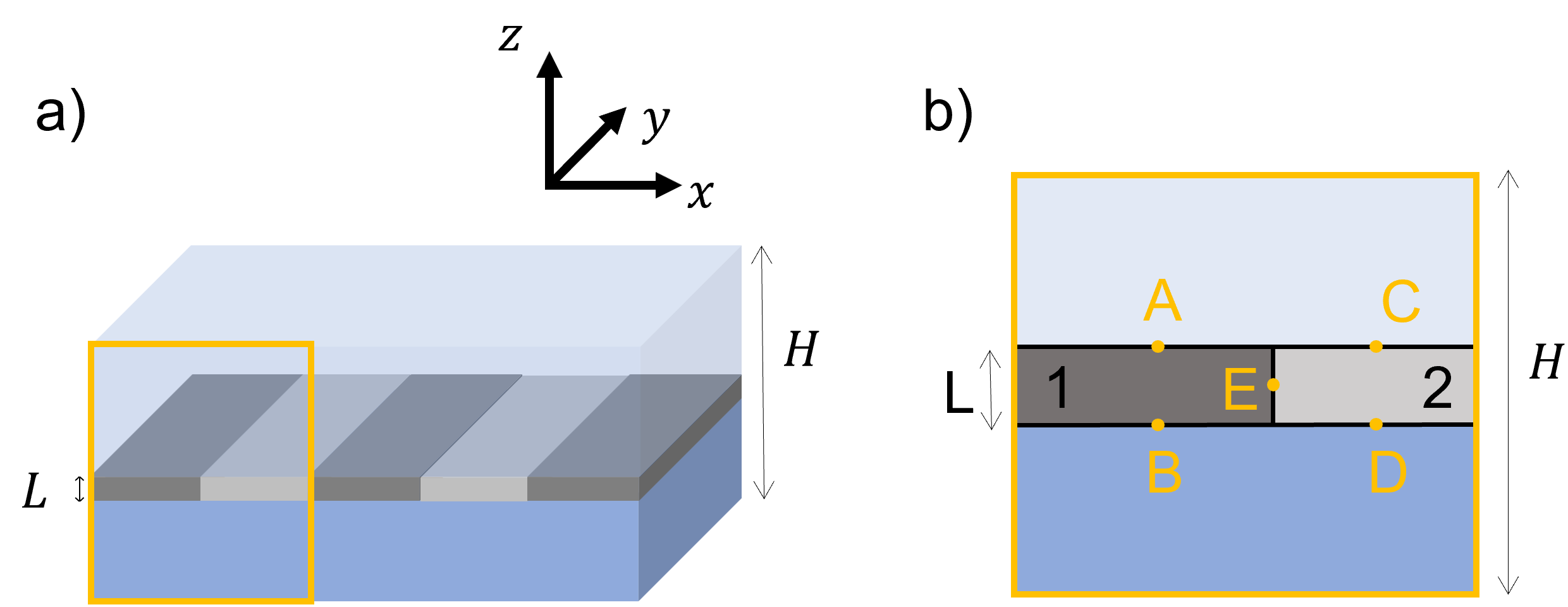}\caption{a) Schematic view of horizontal heterostructure. The yellow box of height H includes the ribbons of 2D materials of thickness $L$  and a part of the substrate and incident medium. b) Close-up view of the yellow box with the different interfaces labelled from A to E.\label{fig:horizontal-heterostructures}}
\par\end{centering}
\end{figure}

In the case of 2D materials, particular care must be taken owing to their
extremely small thicknesses. We first consider the displacement field $\mathbf{D}$  and the electric field $\mathbf{E}$ in a unit cell composed of two distinct materials (fig.~\ref{fig:horizontal-heterostructures}a, yellow rectangle). The incident medium and substrate have large thicknesses compared to the thickness of the 2D ribbon $L$. The total thickness of the system, noted $H$, verifies the condition $H\ll\lambda$.

If the electric field is polarized along the ribbon and parallel to the interface (i.e. along the $y$-axis), it is conserved at the interfaces (namely at points \(A\), \(B\), \(C\), \(D\) and \(E\) depicted in fig. \ref{fig:horizontal-heterostructures}b). Therefore, as the wavelength of the field is much larger than $H$, the electric field is uniform over the whole structure, which is the condition to apply the parallel capacitors equation eq. (S9) (see \cite{SupplementalMaterial}). Accordingly, the effective susceptibility of the layer is 
\begin{equation}
\chi_{S,eff,2D}^{y}=\sum f_{i}\chi_{S,i}^{y}.
\label{eq:ribbon_chi_y}
\end{equation}
similarly to eq. (\ref{eq:ribbon-thick-chi}) with the subscript $2D$ indicating that this equation is valid for 2D materials.

When the electric field is polarized across the ribbon in the material plane ($x$-axis), the electric field ($E^{\parallel}$) is conserved at the interface at points \(A\), \(B\), \(C\) and \(D\). At point \(E\), the displacement field normal to the surface is conserved and the electric field is discontinuous. Therefore, neither the electric field nor the displacement field are constant over the whole structure. The formal conditions to apply strictly the parallel or series capacitors equations are then not fulfilled. 

Nonetheless, as the thickness $L$ of each ribbon is much smaller than its width, one can consider that the electric field does not vary between \(A\) and \(B\) (or between \(C\) and \(D\)). Only close to the interface between the ribbons does the field vary truly.  Consequently, in a first approximation, we consider the electric field constant over the two ribbons, neglecting the variation in the small volume around the interface between the two ribbons. The parallel capacitors equation eq. (S9) applies and:
\begin{equation}
\chi_{S,eff,2D}^{x}=\sum f_{i}\chi_{S,i}^{x},
\label{eq:ribbon_chi_x}
\end{equation}
which gives the same expression than the effective susceptibility in the $y$ polarization (if the 2D materials are isotropic in the plane). The 2D heterostructure is then isotropic in the 2D plane. This is a counter-intuitive result. This conclusion is also different from the one obtained for thick heterostructures (eq. (\ref{eq:ribbon-thick-xi})) that has been used in previous works on 2D materials \cite{Liu2008,Li2020}. We will numerically analyse further these results later in the paper.

Finally, for electric fields across the 2D materials (along the $z$-axis), the displacement field normal to the interface is conserved at the interfaces \(A\), \(B\), \(C\), \(D\) but not \(E\), where the electric field parallel to the interface is conserved. As in the previous case, the displacement field can be approximated as being constant across the ribbons equation, thus (S12) leads to:
\begin{equation}
\xi_{S,eff,2D}^{z}=\sum f_{i}\xi_{S,i}^{z}.
\label{eq:ribbon_xi_z}
\end{equation}

As for the $x$ direction, this is not the same result as for the thick ribbon effective model, which will also be numerically tested later in the paper. 

This effective model for horizontal heterostructures is valid even for ribbon widths of the same order of magnitude as the wavelength. The only necessary condition is that this width needs to be much larger than the layer thickness such that the fields vary slightly in the ribbon.

As a consequence of the in-plane isotropy of the effective model, the optical response of such structured 2D materials at normal incidence does not depend on the polarization except if there are features that cannot be captured by the effective model. For instance, as surface plasmon resonances are phenomena appearing due to the structuring of the material, they cannot be described by the effective layer model. Therefore, comparing the spectra of the effective system to the proper system and inspecting the discrepancies can highlight the plasmonic resonances taking place in the ribbons.

\section{Numerical methods}\label{sec:Methods}

In this section, we describe the reference numerical methods employed to illustrate the range of applicability of the surface susceptibilities and the effective models presented in section \ref{sec:effective_models}. An {\it ab initio} atomistic method is used to determine the susceptibilities of 2D materials and of vertical heterostructures. The optical response is then obtained using the surface susceptibility model of a vertical heterostructure. On the other hand, a classical electrodynamic method is used to determine the absorption of horizontal heterostructures based on the susceptibilities of the individual components. In the last case, {\it ab initio} approaches are not feasible due to the large number of atoms in the unit cell but the horizontal structuring of the materials is fully accounted for, as well as the anisotropy. Those two methods are then complementary in order to verify the relevance of the effective models.

\subsection{Time-dependent density functional theory (TDDFT)}
The surface susceptibility of graphene, hBN and a bilayer graphene-hBN has been calculated using the GPAW implementation of TDDFT~\cite{Enkovaara2010}, within the random-phase approximation. Highly corrugated graphene with a thickness larger than a single layer of flat graphene was also investigated for comparison. The height of the unit cell of graphene and hBN is $1.70$ nm. For the bilayer, the graphene and hBN layers were separated by $0.34$ nm of vacuum, which corresponds to the interlayer distance in graphene and is close to the average interlayer distance of graphene-hBN heterostructures accounting for van der Waals corrections~\cite{Sevilla2021}. In this case, the total height of the cell is $2.0$\,nm. The ground states of graphene, hBN, and the bilayer were calculated using a GGA-PBE functional~\cite{Payne1992}, a k-point grid of $256\times256\times1$ and an energy cut-off of $350$ eV. For the TDDFT calculation, a cut-off energy of 250\,eV was used. Corrugated graphene is modelled using a unit cell containing 50 atoms forming a hill of height $0.25$ nm in a cell of height $2.50$ nm as described in \cite{Dobrik2022}. Its ground state is calculated using the LDA, a $48\times48\times1$ k-point grid and a cut-off energy of 400\,eV. The cut-off energy for the TDDFT calculation is 20\,eV. The GW and BSE calculations are not performed due to computational limitations, which could result in inaccurate optical spectra ~\cite{Onida2002}. However, as we focus here on the effect of the anisotropy on the optical response of heterostructures, we ensure that the same level of approximation is considered when comparing systems. The link between microscopic and macroscopic response functions and effective quantities remains valid for all level of approximations.

\subsection{Rigorous coupled wave analysis }

A homemade code based on the rigorous coupled wave analysis (RCWA) method \cite{Moharam1981} was used. The RCWA method solves Maxwell equations for a series of layers of finite thicknesses with lateral structuring, as horizontal heterostructure. The method was adapted to account for the intrinsic anisotropy of materials in order to accurately model 2D materials \cite{SupplementalMaterial}. The RCWA approach was used for modelling thin films and ribbons, with the dielectric functions obtained by TDDFT.

The thicknesses of the layers representing the 2D materials (structured or not) were arbitrarily fixed to $L=0.34$ nm for monolayer, or $L=0.68$ nm for bilayers. As long as these thicknesses are coherent with the thicknesses used to obtained the effective permittivity the results are independent of this choice ~\cite{Majerus2018}.

\section{Results and discussions}\label{sec:Results}

\begin{figure}
    \centering
    \includegraphics[width=\linewidth, trim = 0cm 8.0cm 0cm 0cm, clip]{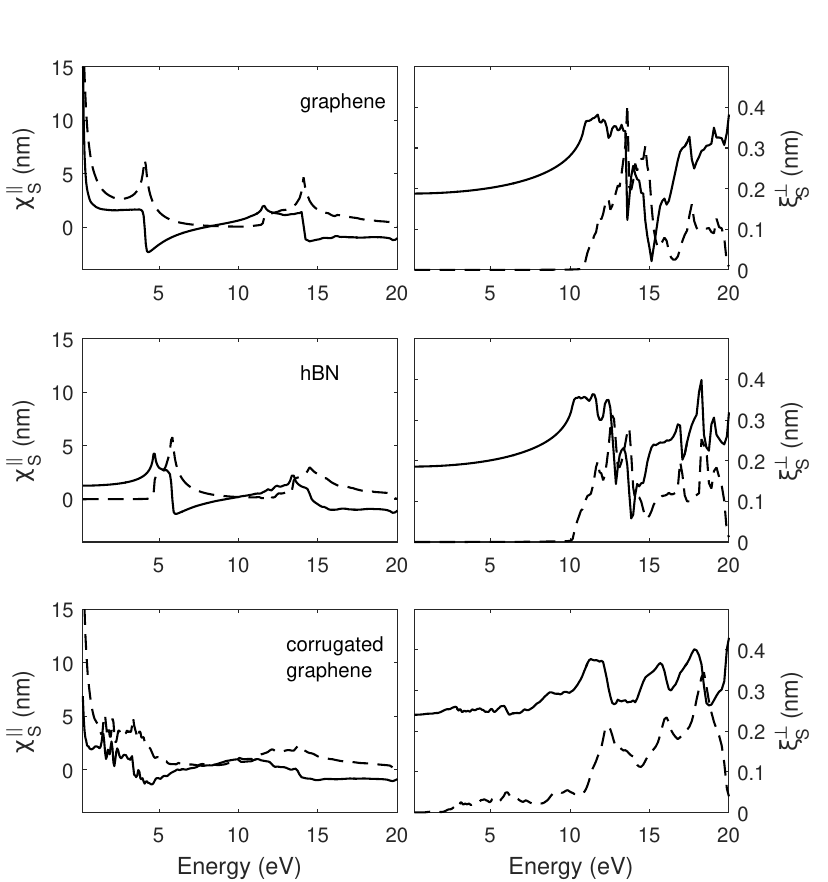}
   \caption{In-plane surface irreducible susceptibility (left) and out-of-plane surface external susceptibility (right) of (from top to bottom) graphene, hBN and corrugated graphene. The solid line is the real part, the dashed line is the imaginary part.}
    \label{Fig:suscept}
\end{figure}

The surface susceptibilities of graphene and hBN are shown on fig.~\ref{Fig:suscept}. It was first verified that $\chi_S^{\parallel}$ and $\xi_s^{\perp}$ are independent of the supercell thickness $L$ (not shown). For the in-plane susceptibility $\chi_S^{\parallel}$ (fig.~\ref{Fig:suscept}, left panels) we observe the $\pi$ and the $\pi + \sigma$ plasmons around respectively $4.5$ and $14$ eV, as expected without the GW and BSE correction \cite{Marinopoulos2004a}. The GW correction tends to blueshift the plasmon energy and the BSE one has the inverse effect. Together they produce a global blueshift of less than $0.5$ eV \cite{Trevisanutto2008a,Chen2011a,Mak2014a,Guilhon2019}. The imaginary part of the out-of-plane susceptibility $\xi_s^{\perp}$, responsible for the absorption, is exactly zero below $10$ eV, fig.~\ref{Fig:suscept}, right panels. For corrugated graphene, the out-of-plane susceptibility is not negligible below $10$ eV, due to the atomic structure extending in the normal direction. This highlights the role of the valence bounds in the normal direction to the out-of-plane response of 2D materials.

The absorption spectra of the systems described by an effective surface polarization were obtained following the method published in our previous papers~\cite{Majerus2018, Majerus2023}. It was adapted to incorporate the surface external susceptibility for the out-of-plane polarization \cite{Majerus2023}. The absorption spectra of the thin films and the ribbons were obtained using the RCWA method.

For all the following absorption calculations of 2D materials, the incident medium is air ($n_a=1$) and the refractive index of the substrate is $n_b=1.5$. The angle of incidence is fixed to 70\textdegree\ in TM polarization to probe the effects of the anisotropy.

\subsection{Vertical heterostructures}
 We consider two types of vertical heterostructures. First, a system with identical layers (graphene) to analyse the limits of the 2D model compared to the thin film model. These structures may be synthesized experimentally up to a few layers using the transfer techniques on CVD-grown graphene for example \cite{Majerus2018a}. Secondly, we test the validity of the effective model for vertical heterostructure (section  \ref{sec:vert_het}) on a graphene-hBN bilayer, whose optical properties have already been reported for in-plane polarizations \cite{Chen2011a,Wang2017,Farmani2017}.

\begin{figure}[htb]
    \centering
    \includegraphics[width=\linewidth, trim = 0cm 0.8cm 0cm 0cm, clip]{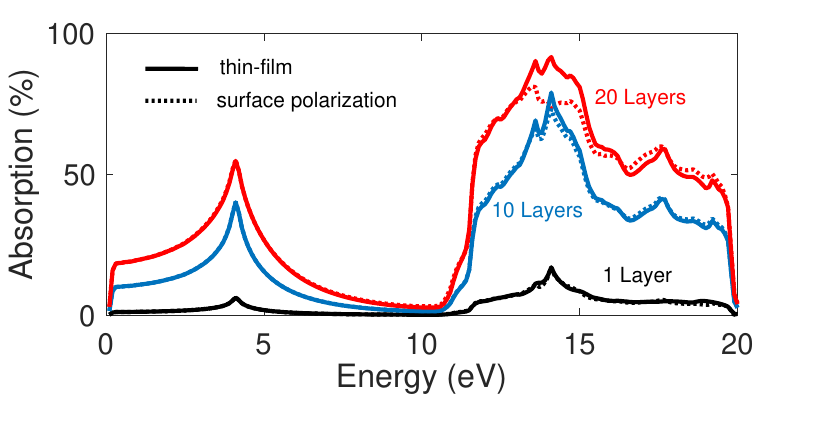}
    \caption{Absorption by a single layer (black), 10 layers (blue) and 20 layers (red) of graphene, considering an incident angle of 70° within the thin-film model (solid line) and the surface polarization model (dotted line).}
    \label{Fig:multilayer}
\end{figure}

In fig.~\ref{Fig:multilayer} the absorption of a single sheet of graphene (black), a multilayer of 10 sheets (blue) and a multilayer of 20 sheets of graphene (red) are calculated using the thin film model (solid lines) and the surface polarization model (dotted lines), showing the robustness of the strictly 2D model, even for 10 layers. For 20 layers, the discrepancies between the models become significant at high energy, in particular around the $\pi+\sigma$ plasmon. In this case, the wavelength of the electromagnetic wave inside the layer is no longer larger than the thickness $L$, and the phase shift cannot be neglected, which was an assumption for the 2D model. However, this result shows that the 2D model is not restricted to single-layered 2D materials and that few-layered 2D materials and heterostructures can also be modelled as surface polarization. The maximum number of layers is determined by the small-phase shift conditions and thus does not depend only on the thickness of the 2D material but also on its permittivity.

\begin{figure}[htb]
    \centering
    \includegraphics[width=\linewidth, trim = 0cm 13.8cm 0.0cm 1.0cm, clip]{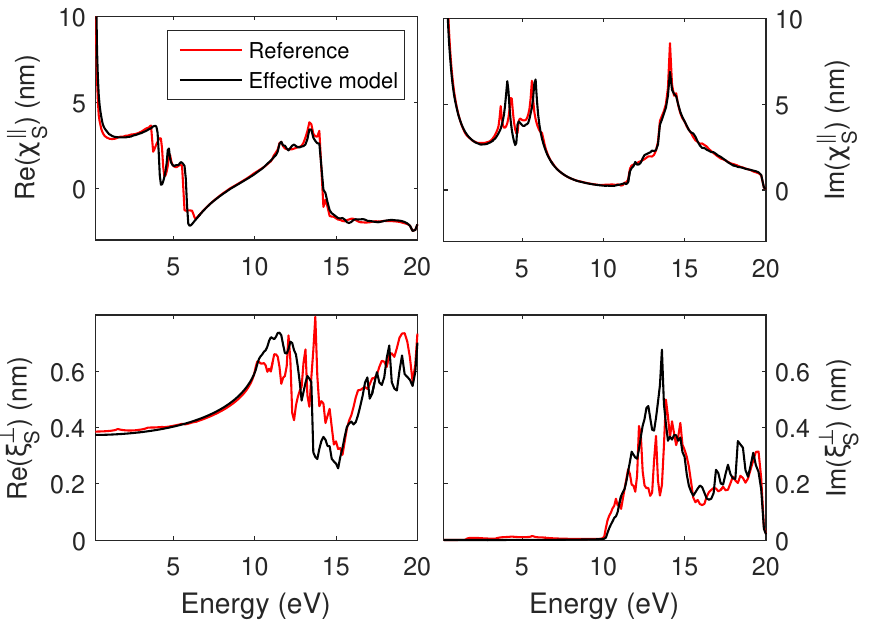}
    \caption{Real (left) and imaginary (right) parts of the surface susceptibilities of a graphene-hBN bilayer heterostructures from the reference model, i.e. TDDFT calculation (red lines) and the effective model (black lines).}
    \label{Fig:graphene-hbn}
\end{figure}

In fig. ~\ref{Fig:graphene-hbn}, the TDDFT surface susceptibilities (the reference) of a graphene-hBN bilayer are compared to the effective susceptibilities of eq.~(\ref{eq:sum_chi_s}) and (\ref{eq:sum_xi_s}), with the single layers susceptibilities also obtained by TDDFT. The effective model replicates well $\chi_S^{\parallel}$ except around $5$ eV which suggests a coupling between the $\pi$-plasmons in each 2D material. For $\xi_S^{\perp}$, the effective model fails to reproduce the TDDFT result above $10$ eV, though the global trend is conserved. This is due to long range electronic interactions that are significant even with large vacuum layers for out-of-plane polarization, as highlighted in \cite{Tancogne-Dejean2015}.

 \subsection{Horizontal heterostructure}
 
The first structure that we consider in this section is made of a 2D pattern alternating between graphene nanoribbons ($15$ nm wide) and vacuum ($5$ nm), leading to a filling factor of $0.75$.  To assess the domain of validity of the thin-ribbon (eqs.~(\ref{eq:ribbon_chi_y})-(\ref{eq:ribbon_xi_z})) and thick-ribbon models (eqs.~(\ref{eq:ribbon-thick-xi}) and (\ref{eq:ribbon-thick-chi})), we compare a single-layer ribbon to a stack of $100$ ribbons, with a total thickness of $35$\,nm.

  \begin{figure}[htb]
    \centering
    \includegraphics[width=\linewidth, trim = 0.3cm 5.3cm 2.5cm 0.5cm, clip]{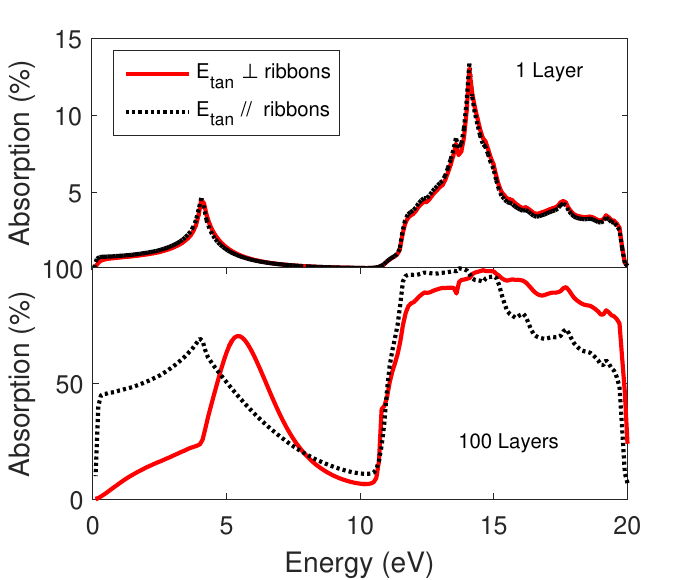}
    \caption{Absorption by ribbons of graphene of 15 nm width with a filling fraction of 0.75 calculated using the RCWA for one layer (top) and one hundred layers (right). The incident angle is 70° and the tangential component of the electric field is either perpendicular (solid red) or parallel to the ribbons (dotted black).}
    \label{Fig:isotropy_ribbons}
\end{figure}

Using the RCWA method, the absorption of the graphene nanoribbons is obtained for two polarizations of the light with an incident angle of $70$°, namely for the component of the electric field in the surface plane $E_{tan}$ either parallel or perpendicular to the ribbons (fig. \ref{Fig:isotropy_ribbons}). To the best of our knowledge, this is the first investigation of 2D materials nanoribbons fully accounting for the intrinsic anisotropy of the 2D material. While for a large thickness (100 layers), the absorption depends on the polarisation, the aligned 2D nanoribbons have an isotropic response. To confirm this result, we also have considered an incident light with a different azimuthal angle, for which the tangential component of the electric field is neither parallel nor perpendicular to the ribbons. No modification of the polarization of the transmitted light has been observed, confirming the isotropy of the system. The uniaxial response of the thin horizontal heterostructure that our effective medium model has evidenced is then confirmed by the RCWA calculation, which fully describes the horizontal structure of the system.

\begin{figure}[htb]
    \centering
    \includegraphics[width=\linewidth, trim = 0.3cm 5.3cm 2.5cm 0.5cm, clip]{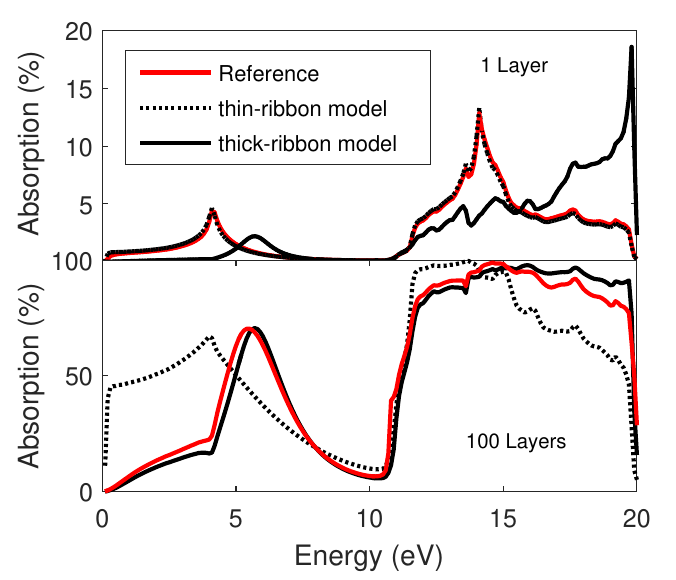}
    \caption{Absorption by ribbons of graphene  with filling factor of 0.75 at an incident angle of 70°, the tangential part of the in-plane electric field is perpendicular to the ribbons, for one layer (top), one hundred layers (bottom). The reference model (RCWA, 15 nm width ribbons) is in red, and the effective models are in dotted (thin-ribbon model) and solid line (thick-ribbon model) black.}
    \label{Fig:thin&thick}
\end{figure}

For thick layers, the effective model (section \ref{sec:effective_models}) predicts that this uniaxial character disappears and, consequently, when $E_{tan}$ is perpendicular to the ribbon, the thick and thin ribbon effective models should differ. In fig.~\ref{Fig:thin&thick}, the reference RCWA simulations are compared to the thin-ribbon and the thick-ribbon effective approaches for a single layer of graphene (top) and for a multilayer of 100 sheets of graphene (bottom). The thin-ribbon model reproduces perfectly the reference results for ribbons of a single sheet of 2D materials while the thick-ribbon model reproduces better the results for the large multilayer. Inversely the thick-ribbon model, which has sometimes been used for 2D materials, gives inaccurate results for single-layer 2D materials \cite{Liu2008,Li2020}.
 
\begin{figure}[htb]
    \centering
    \includegraphics[width=\linewidth, trim = 0cm 0.4cm 0cm 0.4cm, clip]{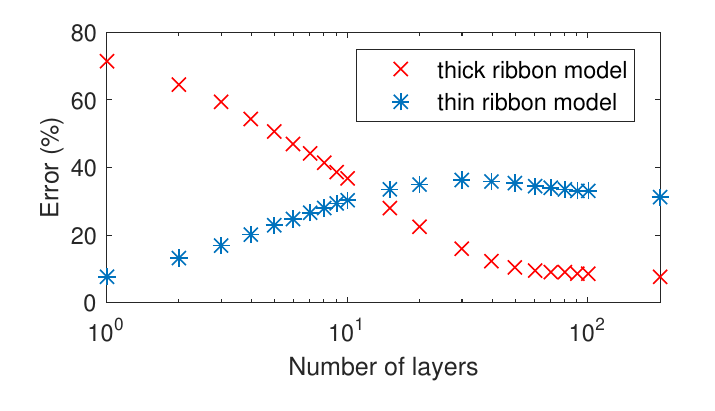}
    \caption{Relative error of the effective models (thin ribbon and thick ribbon) compared to the reference as a function of the number of layer.}
    \label{Fig:ribbon_layers}
\end{figure}

 To better understand the transition between the two models, fig.~\ref{Fig:ribbon_layers} displays the relative error between each effective model and the reference (here the RCWA results) in function of the number of layers. This error is calculated as the normalized area between the curves of absorption $A$: 
 
 \begin{equation}
     Error\,(\%)=100\times \frac{\int\left|A_{eff}-A_{ref}\right|\,dE}{\int A_{ref}\,dE}
 \end{equation}
 with $E$ the incident energy. It confirms that, as shown before, the thin ribbon model is accurate for very few layers but the thick ribbon model works better for several tens of layers. In between, for 3 to 30 layers, the error is above 15\% for both models and a full description of the system is necessary. As mentioned in section \ref{sec:effective_models}, this result is also valid for larger ribbons, as the effective model only depends on the filling factor, as was verified numerically up to a width of $1500$ nm (not shown).

\begin{figure}[htb]
    \centering
    \includegraphics[width=\linewidth, trim = 0cm 0.2cm 0cm 0cm, clip]{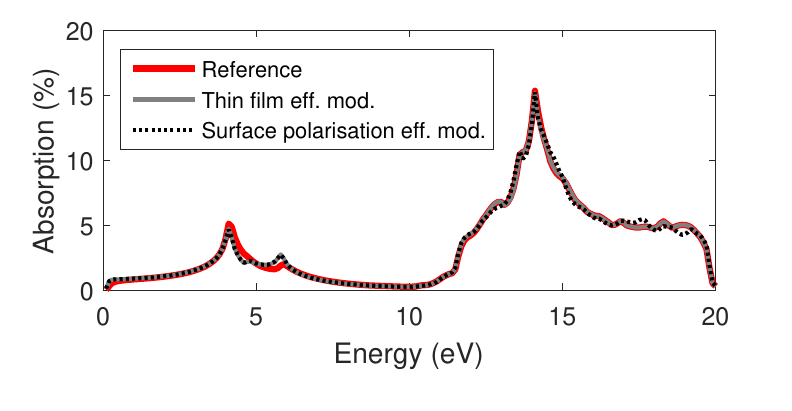}
    \caption{Absorption from alternating ribbons of graphene and hBN, with filling factor of 0.75 and 0.25 respectively, at an incident angle of 70\textdegree, polarized perpendicularly to the ribbons, for three different models: reference model with RCWA (solid red line, width = 15 nm), effective thin film model (solid grey line), effective surface current model (dotted black line). }
    \label{Fig:ribbons}
\end{figure}

We now investigate a lateral repetition of graphene and hBN nanoribbons with of $15$\,nm and $5$\,nm width, respectively. These nanoribbons, have already been produced using CVD and etching device \cite{Levendorf2012,Liu2013}, and may sustain plasmons~\cite{DeAngelis2016,Das2018}, which could be detected by comparison between the results of the effective model and the RCWA calculation. In fig.  \ref{Fig:ribbons}, the three models are compared: the reference model, the effective thin-film model and the effective surface polarization model. The three models agree almost perfectly, except at two specific energies. Around $5$\,eV, the two effective models fail to reproduce the details of the RCWA results because of the plasmonic resonance that cannot be captured by effective models. This discrepancy was not observed in the graphene-air system which suggests that this plasmon originates from a coupling between the $\pi$-plasmon of graphene and that of hBN, similarly to the case of vertical heterostructures. At high energy, above $15$ eV, the RCWA and the effective thin film model results are similar but the surface polarization model slightly differs. In this range, the wavelength is so small that the small phase shift approximation is not valid anymore, invalidating the strictly 2D model.

\section{ Conclusions}

We have proposed an original approach to develop effective models for vertical and horizontal heterostructures of 2D materials, based on the formal link between the microscopic and macroscopic description of the response functions of the materials. The importance of using two different surface susceptibilities, the irreducible susceptibility $\chi_S^{\parallel}$ and the external susceptibility $\xi_S^{\perp}$, defined independently from an arbitrary thickness of the layer, has been highlighted. In particular, it makes it possible to avoid the embarrassing definition of a surface response function that depends on the dielectric properties of the surrounding media. 

It is only recently that experimental optical characterisation highlighted the role of out-of-plane susceptibility ($\xi_S^{\perp}$) for a coherent interpretation of the measurements ~\cite{Xu2021}. It will become even more important with the rapid development of the study of heterostructures. This anisotropy could be taken into account together with the structuration of the material, as in the RCWA methods or in an effective medium approach. In some cases, the out-of-plane response can be neglected as for example, at normal incidence.  Also, below $10$ eV, the imaginary part of $\xi_S^{\perp}$ is negligible for graphene and hBN while the real part is constant. As most of the optical features (absorption, plasmonic excitations,...) depend mainly on the imaginary part of $\xi_S^{\perp}$, the optical spectra are weakly dependent on the out-of-plane response. This justifies a posteriori the use of models without out-of-plane responses in many previous research~\cite{Lobet2016a,Majerus2018a, Madani2013, Zhu2019a}.

For vertical heterostructures, we recovered the well-described effective medium model and expressed it in terms of the susceptibilities. With this effective approach, we reproduce quantitatively calculations from TDDFT although special care must be taken for plasmonic resonance and long-range interactions for out-of-plane polarization. We also demonstrated that the vertical heterostructure effective approach is robust up to tens of layers of graphene (more generally as far as the phase shift of the EM fields is negligible in the heterostructure).

The counter-intuitive uniaxial response of thin alternating nano-ribbons (horizontal heterostructures) is an unexpected outcome of the effective medium approach based on the surface susceptibilities. We successfully confront these predictions with RCWA numerical investigations and illustrate the transition towards an anisotropic behavior as the thickness increases (thick-layer model). This led to the description of an effective model for ribbons of 2D materials different from the effective model for thick ribbons or nanorods. In practice, the validity of the thin-layer model is already questioned for three-layer systems. In both cases, interface excitations, such as surface plasmons, cannot be captured by effective model approaches. As the RCWA method for horizontal heterostructure is numerically very efficient, a full description of the heterostructure rather than an effective medium approach is recommended to avoid questioning the limit of validity. However, the accurate description of systems composed of different 2D materials by a simple homogeneous thin film or surface current presents the obvious advantage of its simplicity and allows to analyse experimental data without numerical effort.

\begin{acknowledgments}
This research used resources of the ``Plateforme Technologique de Calcul Intensif (PTCI)'' (http://www.ptci.unamur.be) located at the University of Namur, Belgium, which is supported by the FNRS-FRFC, the Walloon Region, and the University of Namur (Conventions No. 2.5020.11, GEQ U.G006.15, 1610468, RW/GEQ2016 et U.G011.22). The PTCI is a member of the ``Consortium des Équipements de Calcul Intensif (CÉCI)''.
\end{acknowledgments}

\bibliography{ref_article1.bib}

\end{document}

\title{Anisotropy and effective medium approach in the optical response of 2D material heterostructures : Supplementary information}% Force line breaks with \\

\author{B. Maj\'erus$^{1}$, E. Guillaume$^{1,2,3}$, P. Kockaert$^{4}$ and L. Henrard$^{1}$}

\address{$^{1}$ Laboratoire de physique du solide (LPS) \& Namur Institute of Structured Matters (NISM), University of Namur, 61 rue de Bruxelles, B-5000 Namur, Belgium}
\address{$^{2}$ IMOMEC, IMEC vzw, Wetenschapspark 1, 3590 Diepenbeek, Belgium}
\address{$^{3}$ UHasselt, Institute for Materials Research (IMO-IMOMEC), Agoralaan, 3590 Diepenbeek, Belgium}
\address{$^{4}$ OPERA-photonics, Universit\'e libre de Bruxelles (U.L.B.), 50 Avenue F. D. Roosevelt, CP 194/5, B-1050 Bruxelles, Belgium}

\maketitle

\section{Polarizabilities and susceptibilities} 
 The irreducible polarizability $\alpha^{0}$ (also noted $\chi^{0}$ in some references \cite{Hybertsen1987,Bernadotte2013}) is the response function of non-interacting charges to a perturbing applied potential $\tilde{V}_{app}$, or equivalently, the response function to the total potential $\tilde{V}_{tot}$ (as defined in the main text) \cite{Hybertsen1987,Bernadotte2013}:

\begin{equation}
\rho\left(\mathbf{r}\right)=\int\alpha^{0}\left(\mathbf{r},\mathbf{r}'\right) \tilde{V}_{tot}\left(\mathbf{r}'\right)d\mathbf{r}'
\label{eq:S1}
\end{equation}
where $\rho\left(\mathbf{r}\right)$ is the change in the charge density due to the applied potential. The external polarizability $\alpha$ (also noted $\chi$) is the response function of interacting charges to a perturbing applied potential $\tilde{V}_{app}$:

\begin{equation}
\rho\left(\mathbf{r}\right)=\int\alpha\left(\mathbf{r},\mathbf{r}'\right) \tilde{V}_{app}\left(\mathbf{r}'\right)d\mathbf{r}'
\label{eq:S2}
\end{equation}
The non-interacting polarizability at frequency $\omega$  can be calculated  in the framework of the linear response theory \cite{Hybertsen1987}:
\begin{equation}
\alpha^{0}\left(\mathbf{r},\mathbf{r}',\omega\right)=\sum_{i,j}\left(f_{i}-f_{j}\right)\frac{\phi_{i}\left(\mathbf{r}\right)\phi_{j}^{*}\left(\mathbf{r}\right)\phi_{j}\left(\mathbf{r}'\right)\phi_{i}^{*}\left(\mathbf{r}'\right)}{\omega-E_{j}-E_{i}+i\eta},\label{eq:3.1-polar-wf}
\end{equation}
where $\phi_{i}\left(\mathbf{r}\right)$ is a wave function obtained as a solution of the Kohn-Sham equation, $E_{i}$ is the energy associated to the wave function, $f_{i}$ the Fermi-Dirac distribution at energy $E_{i}$
and $\eta$ a smearing parameter. 
The external polarizability can be calculated using the Dyson-like equation in the random-phase approximation (RPA)~\cite{Yan2011}:
\begin{widetext}
\begin{equation}
%\begin{split}
    \alpha\left(\mathbf{\mathbf{r},\mathbf{\mathbf{r}}'}\right)=\alpha^{0}\left(\mathbf{\mathbf{r},\mathbf{\mathbf{r}}'}\right)\\
    +\int\int\alpha^{0}\left(\mathbf{\mathbf{r},\mathbf{\mathbf{r}}}_{1}\right)\\
    \left(\frac{1}{\left|\mathbf{r}_{1}-\mathbf{r}_{2}\right|}\right)\alpha\left(\mathbf{\mathbf{r}}_{2},\mathbf{\mathbf{r}}'\right)d\mathbf{\mathbf{r}}_{1}d\mathbf{\mathbf{r}}_{2}.
    \label{eq:ext_polar}
%    \end{split}
\end{equation}
\end{widetext}
The microscopic dielectric function $\varepsilon\left(\mathbf{r},\mathbf{r}'\right)$ is often calculated from the polarizabilities and the macroscopic dielectric function is obtained from the averaged values of the dielectric function or its inverse. The method propose in this paper is slightly different but equivalent. In the random-phase approximation (by neglecting the screening from the exchange-correlation term), we define susceptibilities associated to both polarizabilities using coulomb's law on equations (\ref{eq:S1}) and (\ref{eq:S2}) :

\begin{equation}
\chi\left(\mathbf{r},\mathbf{r}'\right)=-\frac{1}{4\pi\varepsilon_{0}}\int\frac{\alpha^{0}\left(\mathbf{r}'',\mathbf{r}'\right)}{\left|\mathbf{r}-\mathbf{r}''\right|}d\mathbf{r}'
\end{equation}

\begin{equation}
\xi\left(\mathbf{r},\mathbf{r}'\right)=-\frac{1}{4\pi\varepsilon_{0}}\int\frac{\alpha\left(\mathbf{r}'',\mathbf{r}'\right)}{\left|\mathbf{r}-\mathbf{r}''\right|}d\mathbf{r}'
\end{equation}
and we verify eq. (4) and (5) of the main paper.

\section{Effective model for multilayers}
The effective model for multilayers is well known and rigorously developed \cite{Bechstedt1986,Mohammadi2005}. However, to extend it to thin ribbon, we propose a simple physical explanation. 
For fields parallel to the layers of a multilayer, the electric field is parallel to the interface. It is then conserved through the interface and consequently, it is constant over all the structure. The effective permittivity is the ratio between the average displacement field and the average electric field. We can thus write
\begin{align}
\varepsilon_{eff}^{\parallel}&=\frac{\left\langle D^{\parallel}\right\rangle }{\varepsilon_{0}\left\langle E^{\parallel}\right\rangle }\\
\varepsilon_{eff}^{\parallel}&=\frac{\sum_{i}f_{i}D_{i}^{\parallel}}{\varepsilon_{0}E^{\parallel}}\\
\varepsilon_{eff}^{\parallel}&=\sum_{i}f_{i}\varepsilon_{i}^{\parallel}\label{S10}
\end{align}
with $\varepsilon_{i}^{\parallel}=\frac{D_{i}^{\parallel}}{\varepsilon_{0} E^{\parallel}}$ the permittivity of the layer  of thickness $d_i$. This equation is called the parallel capacitors equation in the main article.
For fields perpendicular to the layer, the displacement field is conserved at the interface and thus constant over the whole system and we have

\begin{align}
\frac{1}{\varepsilon_{eff}^{\perp}}&=\frac{\varepsilon_{0}\left\langle E^{\perp}\right\rangle }{\left\langle D^{\perp}\right\rangle }\\
\frac{1}{\varepsilon_{eff}^{\perp}}&=\frac{\sum_{i}f_{i}\varepsilon_{0}E_{i}^{\perp}}{D^{\perp}}\\
\frac{1}{\varepsilon_{eff}^{\perp}}&=\sum_{i}f_{i}\frac{1}{\varepsilon_{i}^{\perp}}\label{S14}.
\end{align}

This equation is called the series capacitors equation in the main article.
In brief we observe that if the electric field is constant over the structure, eq. (\ref{S10}) should be used and if the displacement field is conserved, eq. (\ref{S14}) should be used.

\section{Anisotropic RCWA}
\begin{figure}[!h]
    \includegraphics[scale=1]{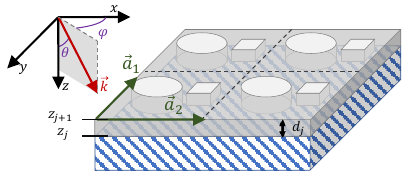}
    \caption{Single layer material (light grey) on a substrate (blue stripes). Periodic repetitions of a circle and square shaped island are depicted.}
    \label{fig:Island_on_sub}
\end{figure}

The Rigorous Coupled Wave Analysis is a method to investigate optical properties of inhomogeneous layered system, first introduced for isotropic media \cite{Moharam1981}. Here, we present the main steps of the methods in the case of anisotropic media, in a similar way than in \cite{Postava2015,Xiang2016,Schlipf2021}. Assuming that each layer of permittivity $\overline{\varepsilon}_h$ is host to a finite number of regions $l$ (e.g. the squares and circle in Fig \ref{fig:Island_on_sub}) of different permittivity ($\overline{\varepsilon}_l$), we can write each component of the permittivity tensor of the layer as:

\begin{equation}
    \varepsilon^{(ij)}\left(\vec{r}\right)=\varepsilon^{(ij)}_h+\sum_l\left(\varepsilon^{(ij)}_l-\varepsilon^{(ij)}_h\right)\Omega^{(l)}\left(\vec{r}\right)
\end{equation}

Where (ij) specifies the component of the permittivity tensor, $\varepsilon_h$ and $\varepsilon_l$ are the permittivities of the host medium and of the $l$-th island, and $\Omega^{(l)}\left(\vec{r}\right)$ is a 2D boolean function delimiting each islands.

Now we expand the permittivity of the whole layer as a 2D Fourier series based upon the Fourier decomposition of the Boolean functions $\Omega^{(l)}$ that define the regions $l$. Since the Fourier expansion is performed on the local functions $\Omega^{(l)}\left(\vec{r}\right)$, we can also compute the Fourier series of any function of the permittivity, e.g. the inverse of any of the component of the tensor:
{\small
\begin{equation}
    \begin{aligned}
        \varepsilon^{(ij)}=\sum_g&\varepsilon^{(ij)}_ge^{i\vec{g}\vdot\vec{r}}\text{ $ $ and $ $ }\frac{1}{\varepsilon^{(ij)}}= \sum_g\eval{\frac{1}{\varepsilon^{ij}}}_ge^{i\vec{g}\vdot\vec{r}}\\
        \text{ $ $ where $ $ }\varepsilon^{(ij)}_g&=\varepsilon^{(ij)}_h\delta_{g,g_0}
        +\sum_l\left(\varepsilon^{(ij)}_l-\varepsilon^{(ij)}_h\right)\Omega_g^{(l)}\\
        \text{ $ $ and $ $ }\eval{\frac{1}{\varepsilon^{(ij)}}}_g&=\frac{1}{\varepsilon^{(ij)}_h}\delta_{g,g_0}+\sum_l\left(\frac{1}{\varepsilon^{(ij)}_l}-\frac{1}{\varepsilon^{(ij)}_h}\right)\Omega^{(l)}_g
    \end{aligned}
\end{equation}}
The Fourier coefficients are indexed along a single index $g$ that runs over all nodes of the 2D reciprocal space, $g_0$ being its origin.

Expressing the permittivity tensor as a Fourier series in Maxwell's equation allows to derive the following set of equations:
\begin{widetext}
{\small
\begin{equation}
    \begin{aligned}
        \dv{E_{z,g'''}}{z}&=-i\sum_{g,g'}\eval{\frac{1}{\varepsilon^{(zz)}}}_{(g'''-g-g')}\hat{\varepsilon}_{g'}\vec{E}_{\parallel,g}\vdot\left(\vec{k}_\parallel+\vec{g}+\vec{g'}\right)
        \text{ $ $ where $ $ } \hat{\varepsilon}_g=
        \begin{pmatrix}
            \eval{\varepsilon^{(xx)}}_{g} & \eval{\varepsilon^{(xy)}}_{g}\\
            \eval{\varepsilon^{(yx)}}_{g} & \eval{\varepsilon^{(yy)}}_{g}
        \end{pmatrix}\\
        \dv{H_{z,g}}{z}&=-i\vec{H}_{\parallel,g}\vdot\left(\vec{k_\parallel}+\vec{g}\right)\\
        \dv{E_{x,g}}{z}&=\frac{i}{\omega\varepsilon_0}\left(k_x+g_x\right)\left[\sum_{g''}\eval{\frac{1}{\varepsilon^{(zz)}}}_{g-g''}
        \left[H_{x,g''}\left(k_y+g_y''\right)-H_{y,g''}\left(k_x+g_x''\right)\right]\right]+i\omega\mu_0H_{y,g}\\
        \dv{E_{y,g}}{z}&=\frac{i}{\omega\varepsilon_0}\left(k_y+g_y\right)\left[\sum_{g''}\eval{\frac{1}{\varepsilon^{(zz)}}}_{g-g''}
        \left[H_{x,g''}\left(k_y+g_y''\right)-H_{y,g''}\left(k_x+g_x''\right)\right]\right]-i\omega\mu_0H_{x,g}\\
        \dv{H_{x,g}}{z}&=\frac{i}{\omega\mu_0}\left(k_x+g_x\right)\left[E_{y,g}\left(k_x+g_x\right)-E_{x,g}\left(k_y+g_y\right)\right]
        -i\omega\varepsilon_0\sum_{g''}\left(\eval{\varepsilon^{(yx)}}_{g-g''}E_{x,g''}+\eval{\varepsilon^{(yy)}}_{g-g''}E_{y,g''}\right)\\
        \dv{H_{y,g}}{z}&=\frac{i}{\omega\mu_0}\left(k_y+g_y\right)\left[E_{y,g}\left(k_x+g_x\right)-E_{x,g}\left(k_y+g_y\right)\right]
        +i\omega\varepsilon_0\sum_{g''}\left(\eval{\varepsilon^{(xx)}}_{g-g''}E_{x,g''}+\eval{\varepsilon^{(xy)}}_{g-g''}E_{y,g''}\right)
    \end{aligned}
\label{eq:3.4-equations-RCWA}
\end{equation}}
\end{widetext}

Note that the non-diagonal components of the tensor enhance the coupling between modes (\textit{intrinsic} anisotropy) in the last two equations, and that unsurprisingly we can see that coupling terms (\textit{structural} anisotropy) are involved irrespective of an isotropic or anisotropic material.

The set of equation shown above can be used to calculate the field at any given depth in a given layer. A stacking of such layers can be described through the Transfer Matrix formalism.

\bibliography{ref_article2}% Produces the bibliography via BibTeX.